\documentstyle[12pt]{article}

\title {Strange Cases from the Files of Astronomical Sociology
\thanks{To be published in {\em Astronomy} magazine,  1993.}}

\author {Kevin Krisciunas}
\date { }

\begin{document}
\maketitle

\vspace{-1 cm}

\begin {center}
   Joint Astronomy Centre \\
   660 N. A'ohoku Place \\
   University Park \\
   Hilo, Hawaii 96720 USA
\end {center}

\vspace{5 mm}

One of the rapidly growing social science fields is that
of astronomical sociology.  Astronomers (and non-astronomers
alike) keep wondering why we are the way we are and what
motivates us.

That astronomers are considered unusual is related, I
think, to our small numbers.  There are some 5900 members of
the American Astronomical Society.  Compare this with the 1.7
million registered nurses and 2.4 million truck drivers in
the United States.  Prior to the middle of the twentieth
century astronomers were a rare breed indeed.  Most
importantly, it does not help our present reputation that
some very unusual stories are associated with astronomers of
the past.

George Joachim Rheticus (1514--1574), who arranged for
the publication of Copernicus' great book, was unable to use
his own surname because his father had been beheaded for
sorcery.\footnotemark[1]

The great pre-telescopic astronomer Tycho Brahe (1546--1601) was
a Danish nobleman and close friend of King Frederick II.   Tycho's
island observatory at Hven cost the equivalent of 30 percent
of the annual revenues of the Danish crown, or about \$5
billion in today's money.\footnotemark[2]  But, on the whole it was well
spent, for Tycho increased the positional accuracy of stellar
positions by a factor of 10, and his observations of Mars led
to Kepler's Laws and Newton's Law of Gravity.

Tycho once had an elk, which he sent as a present to his
mentor, the Landgrave of Hesse-Cassel.  But en route the elk
walked up the steps to a manor house, where it drank a great
quantity of beer.  On the way down the stairs it broke a leg
and died.\footnotemark[3]

It is well-known that Tycho lost the bridge of his nose
in a duel with his third cousin, and that he wore a
prosthetic nose piece, supposedly made of pure gold.  It is
less well-known that Tycho was disinterred in 1901 by some
Czech scholars, who were investigating, among other things,
the composition of his nose piece, which was found to contain
gold, silver, and copper.\footnotemark[4]

The first man to hold an academic chair in science in colonial America
was fired in 1712 for drunkeness and consorting with an ``idle
hussy".\footnotemark[5]

The French astronomer Guillaume Le Gentil (1725--1792) made a valiant
effort to observe the transits of Venus across the disk of the Sun.
These transits took place in 1761 and 1769.  The first transit found
him stuck in the middle of the Indian Ocean, unable to make any
useful observations.  Not to be dissuaded, Le Gentil arrived in India,
built an observatory at a place called Pondicherry, and waited 8 years
for the next transit, which would occur on June 4, 1769.  The weather was
clear for the month prior to the transit, but it clouded up on transit day,
only to clear immediately after the long-awaited event.
Le Gentil then contracted dysentery and remained bedridden for
nine months.  He booked passage home aboard a Spanish warship that was
demasted in a hurricane off the Cape of Good Hope and blown off course
north of the Azores before finally limping into port at Cadiz.  Le Gentil
crossed the Pyrenees on foot and returned to France after an absence of
11 and one-half years, only to learn that he had been declared dead, his
estate looted, and its remains divided up among his heirs and creditors,
after which he gave up astronomy.\footnotemark[6]

The Russian astronomer Wilhelm Struve (1793--1864) produced 272
astronomical works and 18 children.  Otto Struve (1897--1963) produced
907 works and had zero children.\footnotemark[7]

The French astronomy popularizer Camille Flammarion
(1842--1925) was the love object of a French countess who died
at a young age of tuberculosis.  They never even met, but the
young woman made an unusual request to her doctor  --- that when
she died he would cut a large piece of skin from her back and
bring it to Flammarion with the request that he have it
tanned, and that it be used to bind a copy of his next book.
And so it happened.  Flammarion's first copy of {\em Terres du
Ciel} was so bound, with an inscription in gold in the front
cover: Pious fulfillment of an anonymous wish/ Binding in
human skin (woman) 1882.\footnotemark[8]

The most egotistical work in the history of astronomy is
the biography by W. L. Webb of the American astronomer T. J.
J. See (1866--1962), known to some as ``The Sage of Mare
Island."  This book, chronicling the ``unparalleled"
discoveries of See, is in fact an autobiography written in
the third person, using a pseudonym.\footnotemark[9]

George Ellery Hale (1868--1938) was the founder of Yerkes
Observatory, then of Mt. Wilson Observatory.  Beginning at age
42 Hale received regular visits from an elf, who advised him
on numerous matters, including the administration of Mt. Wilson
and the planning for Palomar.\footnotemark[10]

Hale, Wilhelm Struve, Charles Piazzi Smyth (1819--1901),\footnotemark[11]
and this author all spent part of a honeymoon visiting observatories or
doing astronomical site testing.

One of the most brilliant (and most irascible)
astronomers of all time was Fritz Zwicky (1898--1974).  He
made many significant breakthroughs, among them the discovery
of the ``dark matter" permeating the Coma cluster of galaxies.

In the January 19, 1934 edition of the Los Angeles {\em Times }
Zwicky was lampooned in a cartoon entitled ``Be Scientific
with Ol' Doc Dabble," in which it says: ``Cosmic rays are
caused by exploding stars which burn with a fire equal to 100
million suns and then shrivel from 1/2 million mile diameters
to little spheres 14 miles thick."  In the polemical introduction to
Zwicky's {\em Catalogue of Selected Compact Galaxies and of
Post-Eruptive Galaxies} (1971), known simply as ``The Red Book,"
Zwicky quotes the cartoon and comments: ``This, in all
modesty, I claim to be one of the most concise triple
predictions ever made in science."  Why?  Because it
correctly describes the nature of origin of cosmic rays,
supernovae, and the formation of neutron stars.

Zwicky was fond of standing up in seminars to remind the
speaker that Zwicky had solved the particular question many
years before.  He also used to refer to other astronomers at
Mt. Wilson and Palomar as {\em spherical} bastards.  Why
``spherical"?  Because they were bastards any way you look at
them.  One time Zwicky had the night assistant at the 200-
inch fire a bullet out the dome slit in the direction the
telescope was pointing to see if that improved the seeing.
It did not.\footnotemark[12]

Russian-born Harvard astronomer Sergei Gaposchkin (1898--
1984?) was once arguing with a night assistant at McDonald
Observatory about why they could not find a particular
object.  The night assistant suggested that one way or
another the coordinates were incorrect.  Gaposchkin retorted,
``I, Sergei Gaposchkin, am {\em never} wrong!"  And with that he
took one step backward, fell off the observing platform and
broke his arm.\footnotemark[13]  Gaposchkin's curious last publication was
called, ``The 22nd most remarkable star RY Scuti."\footnotemark[14]
It contains, among other things, a reference to a Frank
Sinatra song.

I could go on and on, and in fact I often do.  And in fact
I recently joined a self-help group for people that talk too
much.  It's called On and On Anon.
But to conclude, I often wonder, as my friends suggest,
if the term ``eccentric astronomers" is repetitively redundant.
Do astronomers become eccentric as a result of environmental
factors such as sleep deprivation and jet lag, or do
inherently eccentric people go into this business so as to
have somewhere to fit in, sort of?  Astronomical sociologists
are hard at work trying to answer these questions.

\newpage

\begin{center}
{\bf Footnotes}
\end{center}

1. {\em Concise Dictionary of Scientific Biography}, (New York:
Scribner's), 1981, p. 583.

2. Krisciunas, Kevin, {\em Astronomical Centers of the World},
(Cambridge: Cambridge Univ. Press), 1988, p. 44.

3. Dreyer, J. L. E., {\em Tycho Brahe: A Picture of Scientific Life
and Work in the Sixteenth Century}, (Edinburgh: Adam \& Charles Black),
1890, p. 210.

4. Ashbrook, Joseph, ``Tycho Brahe's nose," {\em Sky and Telescope},
June 1965, p. 353.

5. Yeomans, D. K., ``The shaky beginning of North American astronomy,"
{\em Bull. Amer. Astr. Soc.}, {\bf 11}, 1979, p. 602.

6. Ferris, Timothy, {\em Coming of Age in the Milky Way}, (New York:
William Morrow), 1988, pp. 133-4.

7. Krisciunas, Kevin, ``Otto Struve (1893-1964)," {\em Biographical Memoirs},
{\bf 61}, (Washington, D. C.: National Academy Press), 1992, p. 351.

8. Blumenthal, Walter Hart, {\em An Olio Bookmen's Bedlam of Literary
Oddities}, (New Brunswick, New Jersey: Rutgers Univ. Press), 1955,
p. 85 ff.

9. Ashbrook, Joseph, ``The sage of Mare Island," {\em Sky and Telescope},
October 1962, p. 193.

10. Preston, Richard, {\em First Light: The Search for the Edge of the
Universe}, (New York: New American Library), 1987, p. 37.

11. Sanchez, F., ``Astronomy in the Canary Islands,"{\em Vistas
in Astronomy}, {\bf 28}, 1985, pp. 417-430, on p. 418.  Piazzi Smyth and
his wife spent nearly half of their 113 day honeymoon trip camping out
in cold, rainy conditions at altitude on Tenerife.

12. Preston, {\em op. cit.}, p. 114.

13. The bare bones of this story is found in: Evans, David S., and
Mulholland, J. Derral, {\em Big and Bright: A History of the McDonald
Observatory}, (Austin: Univ. of Texas Press), 1986, p. 93.  I first
heard the story from a Lick Observatory astronomer who was a Harvard
graduate student.

14. {\em Bull. Amer. Astr. Soc.}, {\bf 11}, 1979, p. 402.

\end{document}